\documentclass[ba]{imsart}
\RequirePackage{amsthm,amsmath,amsfonts,amssymb}
\RequirePackage[numbers]{natbib}
\RequirePackage[colorlinks,citecolor=blue,urlcolor=blue,backref=page,backref=page]{hyperref}
\RequirePackage{graphicx}
\usepackage{physics}
\usepackage[utf8]{inputenc}
\usepackage[english]{babel}
\usepackage[T1]{fontenc}
\usepackage{lmodern}
\usepackage{microtype}
\usepackage{booktabs}
\usepackage{bm}
\usepackage{listings}
\usepackage{subfigure}
\usepackage{fancyhdr}
\usepackage{mathtools}
\usepackage{comment}
\usepackage{color}
\usepackage{upgreek}

\pubyear{2024}
\arxiv{XXXX.XXXXX}
\volume{TBA}
\issue{TBA}
\firstpage{1}
\lastpage{1}

\startlocaldefs
\newcommand{\prior}[1]{\Pi(#1)}
\newcommand{\like}[1]{\mathcal{L}(#1)}
\newcommand{\post}[1]{\mathcal{P}(#1)}
\newcommand{\evid}{\mathcal{Z}}
\newcommand{\hyp}[1]{\mathcal{H}_\mathrm{#1}}
\newcommand{\bayesf}[2]{\mathcal{B}^\mathrm{#1}_\mathrm{#2}}
\newcommand{\NIX}{\mathrm{NI}\chi^2}

\endlocaldefs

\begin{document}

\begin{frontmatter}
\title{Hierarchical inference of evidence using posterior samples}
\runtitle{Evidence inference using posterior samples}

\begin{aug}
\author[A]{\fnms{Stefano}~\snm{Rinaldi}\ead[label=e1]{stefano.rinaldi@uni-heidelberg.de}\orcid{0000-0001-5799-4155}},
\author[B]{\fnms{Gabriele}~\snm{Demasi}\ead[label=e2]{gabriele.demasi@unifi.it}\orcid{0009-0009-5320-502X}},
\author[C]{\fnms{Walter}~\snm{Del~Pozzo}\ead[label=e3]{walter.delpozzo@unipi.it}\orcid{0000-0003-3978-2030}}\and
\author[D]{\fnms{Otto~A.}~\snm{Hannuksela}\ead[label=e4]{hannuksela@phy.cuhk.edu.hk}\orcid{0000-0002-3887-7137}}
\address[A]{Institut~für~Theoretische~Astrophysik, ZAH, Universität~Heidelberg, Albert-Ueberle-Stra{\ss}e~2, 69120 Heidelberg, Germany\printead[presep={,\ }]{e1}}
\address[B]{Dipartimento di Fisica e Astronomia, Università degli Studi di Firenze, Via Sansone 1, 50019 Sesto Fiorentino, Italy\printead[presep={,\ }]{e2}}
\address[C]{Dipartimento di Fisica ``E. Fermi'', Università di Pisa, Largo Bruno Pontecorvo 3, 56127 Pisa, Italy\printead[presep={,\ }]{e3}}
\address[D]{Department of Physics, The Chinese University of Hong Kong, Shatin, New Territories, Hong Kong\printead[presep={,\ }]{e4}}
\runauthor{S. Rinaldi et al.}
\end{aug}

\begin{abstract}
The Bayesian evidence, crucial ingredient for model selection, is arguably the most important quantity in Bayesian data analysis: at the same time, however, it is also one of the most difficult to compute.
In this paper we present a hierarchical method that leverages on a multivariate normalised approximant for the posterior probability density to infer the evidence for a model in a hierarchical fashion using a set of posterior samples drawn using an arbitrary sampling scheme.
\end{abstract}



\end{frontmatter}

\section{Introduction}
Quantitatively selecting the best description of some observations among different competing models is one of the most powerful possibilities offered by the Bayesian approach to statistics. This is done making use of the so-called \emph{Bayesian evidence}, the probability associated with a specific model conditioned on the available observations \citep{sivia:2006}. This number, despite not being self-consistent \emph{per se} and thus not capable of assessing in an absolute sense the goodness of a model, can be used to compute the \emph{relative} probability of two models and, therefore, to provide a ranking statistics.

Throughout the paper, we will make use of the following notation:
\begin{itemize}
\item $\prior{x}$ is the prior probability density;
\item $\like{x}$ is the likelihood function;
\item $\evid$ denotes the evidence, which acts as a normalisation constant for the posterior;
\item $\post{x}$ is the normalised posterior probability density.
\end{itemize}
These quantities are related by the Bayes' theorem:
\begin{equation}
\evid\post{x} = \like{x}\prior{x}\,.
\end{equation}

The evidence $\evid$, also called \emph{marginal likelihood}, is the integral over all the parameter space $\mathcal{X}$ of the unnormalised posterior distribution:
\begin{equation}\label{eq:evidence}
\evid = \int_\mathcal{X} \like{x}\prior{x}\dd x\,.
\end{equation}
This integral requires the evaluation of a (often) highly variable likelihood function over a potentially high-dimensional parameter space. These two facts make this integral numerically challenging to solve in a reliable way for a large number of cases. Nonetheless, the scientific reward promised by the Bayesian evidence pushed the development of a variety of methods specifically designed to compute this quantity. Among the most widely used in astronomy, we find thermodynamic integration \citep{hobson:2003} and nested sampling (NS) \citep{skilling:2006}. These schemes, however, are computationally expensive and take a long time to converge for non-trivial problems, especially in all these situations where there is a degeneracy among the parameters.

Another possibility that has been explored is the evaluation of the evidence integral in Eq.~\eqref{eq:evidence} making use of samples drawn from the posterior distribution $\post{x}$. An appealing possibility is to use the so-called \emph{harmonic mean estimator} \citep{newton:1994}, which is defined as
\begin{equation}\label{eq:harmonicmean}
\evid \simeq \qty(\frac{1}{n}\sum_{i=1}^n \frac{1}{\like{x_i}})^{-1}\qc x_i\sim\post{x}\,.
\end{equation}
This is the Monte Carlo estimation of
\begin{equation}
1 = \int_\mathcal{X} \prior{x}\dd x = \evid\int_\mathcal{X} \frac{1}{\like{x}}\frac{\like{x}\prior{x}}{\evid} \dd x\,,
\end{equation}
which can be evaluated making use of the importance sampling technique.

As initially noted by \citet{neal:1994} and subsequently confirmed by other works \citep{clyde:2007,friel:2012}, the harmonic mean estimator is unstable with respect to the prior choice and its variance is infinite. A workaround for this instability was proposed by \citet{gelfand:1994} introducing the \emph{re-targeted} harmonic mean estimator:
\begin{equation}\label{eq:retargeted}
\evid \simeq \qty(\frac{1}{n}\sum_{i=1}^n \frac{\varphi(x_i)}{\like{x_i}\prior{x_i}})^{-1}\qc x_i\sim\post{x}\,.
\end{equation} 
Here $\varphi(x)$ denotes a probability density function meant to stabilise the harmonic mean estimator. $\varphi(x)$ must satisfy the following two conditions:
\begin{enumerate}
\item $\varphi(x)$ must be normalised;
\item The support of $\varphi(x)$ must be entirely contained within the support of $\post{x}$.
\end{enumerate}
Choosing an appropriate functional form for $\varphi(x)$ is a challenging task, especially for high-dimensional problems \citep{chib:1995}. The optimal choice for $\varphi(x)$ would be $\post{x}$ itself, but it would require the knowledge of the normalisation constant $\evid$. Other possibilities have been explored in literature, such as the multivariate Gaussian distribution \citep{gelfand:1994}, indicator functions \citep{robert:2009} or machine learning based methods \citep{caldwell:2018,mcewen:2021,spuriomancini:2023,polanska:2023}. 
Beyond the ones mentioned above, which inspired this work, several other methods to compute the evidence using posterior samples have been proposed, such as bridge sampling \citep{meng:1996,gelman:1998}, dual importance sampling \citep{lee:2016}, Laplace approximation \citep{tierney:1986} and normalising flows \citep{srinivasan:2024}.

In this paper we present an approach designed not to compute the evidence integral but rather to \emph{infer} its value making use of a set of posterior samples from the posterior distribution in a hierarchical framework that first approximates $\post{x}$ with a flexible model and subsequently leverages on the fact that the approximant is normalised to obtain a probability distribution for the evidence $\evid$. This approach, inspired from hierarchical population studies, to the best of the authors' knowledge has never been proposed.
The paper is organised as follows: Section~\ref{sec:methods}, after a brief overview of non-parametric methods, gives the details of the proposed  hierarchical approach. In Section~\ref{sec:examples} we apply our method to some increasingly complex problems, whereas Section~\ref{sec:bayesfactor} presents the application of this scheme to a model selection problem.

\section{Methods}\label{sec:methods}
In this section we give a brief overview of Bayesian non-parametric methods and we introduce the proposed hierarchical scheme to infer the evidence value using a set of posterior samples.
In what follows, we will denote sets with boldface symbols, e.g. $\mathbf{x} = \{x_1,\ldots,x_n\}$, and sets of sets with capital boldface symbols, $\mathbf{Y} = \{\mathbf{y}_1,\ldots,\mathbf{y}_n\}$.

\subsection{Non-parametric methods: DPGMM and (H)DPGMM}
In this paper we propose an approach to infer the value of $\evid$, rather than computing it, leveraging on a non-parametric approximant for $\post{x}$. 

Bayesian non-parametric methods are powerful tools to reconstruct arbitrary probability densities without being committal to any specific functional form \citep{gelman:2013}. In particular, they are useful in all those situations when a parametric model\footnote{To be exact, also non-parametric models have parameters: the difference lies in the fact that, for these, the number of parameter is infinite and and they cannot be directly interpreted in terms of the underlying processes that generated the observations.} is not available because the underlying processes are not understood well enough or because one is simply aiming at getting an effective representation of the inferred probability distribution. Among the different non-parametric methods that are available in literature, we will focus our attention on the Dirichlet process Gaussian mixture model \citep{escobar:1995,rasmussen:2000}, or DPGMM for short: this model, which is an infinite weighted sum of Gaussian distribution, is able to approximate arbitrary probability densities in multiple dimensions \citep{nguyen:2020}:
\begin{equation}\label{eq:dpgmm}
p(x) \simeq \sum_{i=1}^{\infty} w_i \mathcal{N}(x|\mu_i,\sigma_i) \equiv \mathrm{DPGMM}(x|\theta)\,.
\end{equation}
The (countably infinite) parameters of this model are the weights $\mathbf{w} = \{w_1,w_2,\ldots\}$, the mean vectors $\boldsymbol \mu = \{\mu_1,\mu_2,\ldots\}$ and the covariance matrices $\boldsymbol \sigma = \{\sigma_1,\sigma_2,\ldots\}$: we will denote these parameters collectively as $\theta = \{\mathbf{w},\boldsymbol\mu,\boldsymbol\sigma\}$. Samples $\theta_j$ can be drawn from the posterior distribution conditioned on a set of posterior samples $\mathbf{x} = \{x_1,\ldots x_n\}$ drawn from $p(x)$:
\begin{equation}
\theta_j \sim p(\theta|\mathbf{x})\,.
\end{equation}

In \citet{rinaldi:2022:hdpgmm} we introduce \emph{a hierarchy of Dirichlet process Gaussian mixture models}, or (H)DPGMM, a generalisation of the DPGMM to the case in which one does not have direct access to the individual samples $x_i$ but rather on a posterior distribution $p_i(x_i)$ represented by a set of posterior samples $\mathbf{y}_i = \{y_{i,1},\ldots,y_{i,m}\}$:
\begin{equation}
y_{i,j} \sim p_i(x_i)\,.
\end{equation}
In this model, both every individual posterior distribution $p_i(x_i)$ and the hyper-distribution $p(x)$ is approximated by a DPGMM: following the convention of \citet{rinaldi:2022:hdpgmm}, we will call \emph{outer} DPGMM the one approximating $p(x)$ and \emph{inner} the ones describing $p_i(x_i)$. Samples for the parameters of the outer DPGMM, denoted with $\Theta_j$, can be drawn from the marginalised posterior distribution:
\begin{equation}\label{eq:hdpgmm}
\Theta_j \sim p(\Theta|\mathbf{Y}) = \int p(\Theta|\mathbf{x})\prod_{i=1}^np(x_i|\theta_i) p(\theta_i|\mathbf{y}_i) \dd \mathbf{x}\dd \boldsymbol \theta\,.
\end{equation} 
Here we denoted $\mathbf{Y} = \{\mathbf{y}_1,\ldots\mathbf{y}_n\}$ and $\boldsymbol \theta = \{\theta_1,\ldots\theta_n\}$. Both DPGMM and (H)DPGMM can be efficiently explored using the collapsed Gibbs sampling scheme \citep{liu:1994}: in particular, in this paper we make use of \textsc{figaro} \citep{rinaldi:2022:figaro}, a \textsc{Python} implementation\footnote{Publicly available at \url{https://github.com/sterinaldi/FIGARO} and via \href{https://pypi.org/project/figaro/}{\texttt{pip}}.} designed to reconstruct arbitrary probability densities using either the DPGMM or (H)DPGMM.

\begin{figure*}
    \centering
    \centering
    \subfigure[Linear scale]{
        \includegraphics[width = 0.47\columnwidth]{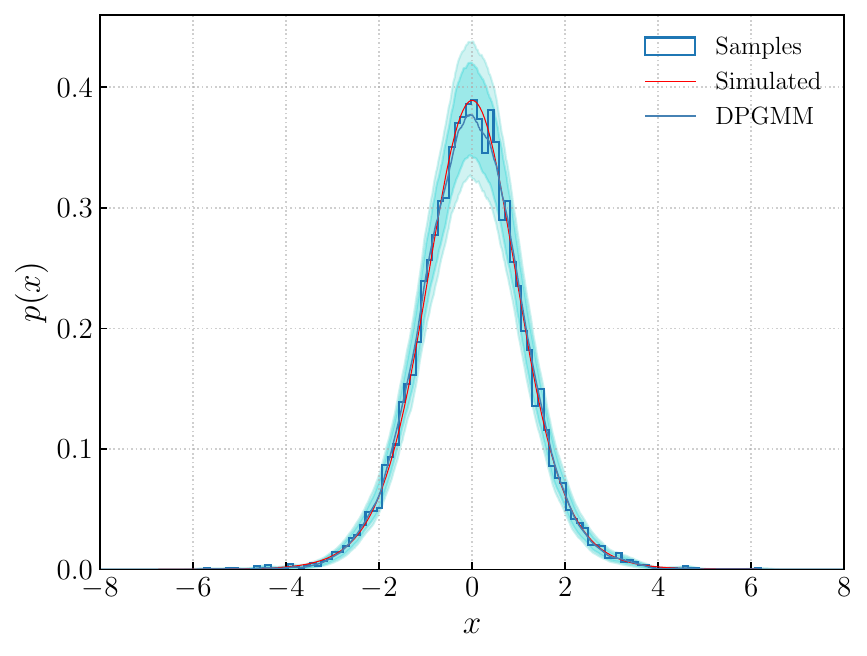}
        \label{fig:log_student-t}
    }
    \subfigure[Logarithmic scale]{
        \includegraphics[width = 0.47\columnwidth]{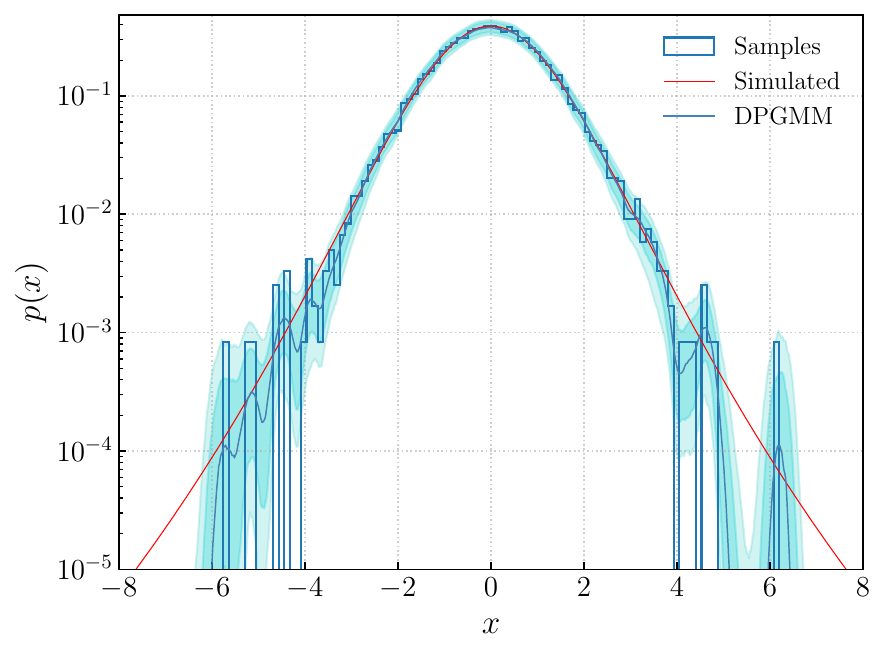}
        \label{fig:lin_student-t}
    }
    \caption{DPGMM reconstruction (blue line, shaded regions correspond to 68\% and 90\% credible regions) of a Student-t distribution with 10 degrees of freedom (red line) using $10^4$ samples (histogram). The heavy tails of the Student-t distribution are not properly sampled, thus the DPGMM reconstruction is concentrated around the available samples.}
    \label{fig:student-t}
\end{figure*}

The DPGMM is an accurate approximant for the bulk of the distribution: however, the agnosticism and flexibility of this model comes with the cost that, since all the information on the underlying distribution comes from the samples themselves, the approximation in Eq.~\eqref{eq:dpgmm} does not hold in all these regions of the parameter space where the sample density is low. This is the case of the tails of the distribution, where the approximant will be peaked around the few samples available. We give an example of this in Figure~\ref{fig:student-t}, where we show the DPGMM reconstruction of a Student-t distribution with 10 degrees of freedom using $10^4$ samples. This feature makes the DPGMM reconstruction not suitable as a stabilising function $\varphi(x)$ for the re-targeted harmonic mean estimator.

\subsection{Hierarchical inference of $\evid$}\label{sec:hierinference}
In principle, having at hand a perfect approximant $\mathcal{A}(x)$ for the posterior probability density $\post{x}$, a single evaluation of the ratio 
\begin{equation}
\hat\evid_i = \evid(x_i) = \frac{\like{x_i}\prior{x_i}}{\mathcal{A}(x_i)}
\end{equation}
would be sufficient to obtain $\evid$, since every single value $\hat\evid_i$ would be equal to the actual evidence. In this respect, each sample $x_i$ drawn from $\post{x}$ can be transformed into a sample $\hat\evid_i$ drawn from a delta distribution:
\begin{equation}
\hat\evid_i \sim p(\hat\evid|\evid) = \delta(\hat\evid-\evid)\,.
\end{equation}
The DPGMM reconstruction presented in the previous section, despite not being arbitrarily precise, has the advantage of providing an approximant for the posterior probability density which is already normalised and distributed around $\post{x}$. However, this reconstruction comes with an uncertainty budget -- see, for example, the shaded regions in Figure~\ref{fig:lin_student-t}. Moreover, there is no reason to expect that any point estimate for $\post{x}$ obtained from the DPGMM reconstruction, such as the median value of $\mathrm{DPGMM}(x|\theta)$ taken over all the realisations $\theta_i$ (blue line in Figure~\ref{fig:student-t}), will match $\post{x}$ perfectly.\footnote{This is due to the intrinsic fluctuations induced by the finite number of samples used in the reconstruction.} However, marginalising over the parameters $\theta$ of the DPGMM reconstruction gives a probability density for the specific value of $\hat\evid_i$ (the one corresponding to the sample $x_i$) conditioned on the samples $\mathbf{x}$:
\begin{equation}
p(\hat\evid_i|\mathbf{x}) = \int p(\hat\evid_i|x_i,\theta)p(\theta|\mathbf{x})\dd\theta= \int \delta\qty(\hat\evid_i - \frac{\like{x_i}\prior{x_i}}{\mathrm{DPGMM}(x_i|\theta)})p(\theta|\mathbf{x})\dd\theta\,.
\end{equation}
In practice, samples from the distribution $p(\hat\evid_i|\mathbf{x})$ are obtained by taking all the available realisations of $\theta_j$ drawn from $p(\theta|\mathbf{x})$ and applying a change of variable:
\begin{equation}\label{eq:variablechange}
\evid_{i,j} = \frac{\like{x_i}\prior{x_i}}{\mathrm{DPGMM}(x_i|\theta_j)}\qc\theta_j \sim p(\theta|\mathbf{x})\,.
\end{equation}

The individual distributions $p(\hat\evid_i|\mathbf{x})$ will be scattered around the true evidence value, due to the finite number of samples drawn from $\post{x}$ used to obtain the approximant: local under/over-densities of such samples result in different expected values for $p(\hat\evid_i|\mathbf{x})$ for each $\hat\evid_i$. Making use of a hierarchical inference framework allows us to account for these stochastic fluctuations, eventually resulting in a probability density for $\evid$. 
This unknown probability density can be approximated with a DPGMM:
\begin{equation}
p(\evid) \simeq \mathrm{DPGMM}(\evid|\Theta)\,.
\end{equation}
The parameters $\Theta$, in turn, are distributed according to the following probability density:
\begin{equation}\label{eq:post_theta_1}
p(\Theta|\mathbf{x}) = \int p(\Theta|\boldsymbol{\hat\evid})p(\boldsymbol{\hat\evid}|\mathbf{x})\dd \boldsymbol{\hat\evid} = \int p(\Theta|\boldsymbol{\hat\evid})p(\boldsymbol{\hat\evid}|\mathbf{x},\theta)p(\theta|\mathbf{x})\dd\boldsymbol{\hat\evid}\dd\theta\,.
\end{equation}
Here we denoted with $\boldsymbol{\hat\evid} = \{\hat\evid_1,\ldots,\hat\evid_n\}$.
Again, each distribution $p(\hat\evid_i|\mathbf{x})$ can be approximated with a DPGMM: expanding the term $p(\boldsymbol{\hat\evid}|\mathbf{x},\theta)$ in Eq.~\eqref{eq:post_theta_1} we get:
\begin{equation}\label{eq:post_theta_2}
p(\Theta|\mathbf{x}) = \int p(\Theta|\boldsymbol{\hat\evid})\prod_ip(\hat\evid_i|\upvartheta_i)p(\upvartheta_i|x_i,\theta) p(\theta|\mathbf{x})\dd \boldsymbol{\hat\evid}\dd\theta\,.
\end{equation}
In this equation, $\upvartheta_i$ denotes the parameters of the DPGMM used to approximate $p(\hat\evid_i|\mathbf{x})$. This equation, besides the marginalisation over $\theta$ that is simply used to obtain a posterior probability density for $p(\hat\evid_i|\mathbf{x})$, has the same form of Eq.~\eqref{eq:hdpgmm}: therefore, we can apply (H)DPGMM to hierarchically reconstruct a posterior probability density for $\evid$ given a set of samples $\mathbf{x}$. The steps to follow can be summarised as:
\begin{enumerate}
\item Draw a number of $\theta_j$ samples from $p(\theta_j|\mathbf{x})$. This gives an approximant for $\post{x}$;
\item For each sample $x_i$ in a subset located in the bulk of $\post{x}$, e.g. a number of randomly selected samples among the 50\% with the largest value of $\like{x}\prior{x}$, compute $\boldsymbol{\hat\evid}_i = \{\hat\evid_{i,1},\ldots,\hat\evid_{i,m}\}$ applying Eq.~\eqref{eq:variablechange} to all $\theta_j$;
\item Reconstruct each probability density $p(\hat\evid_i|\mathbf{x})$ with a DPGMM using $\boldsymbol{\hat\evid}_i$;
\item Combine all the DPGMM reconstructions of $p(\hat\evid_i|\mathbf{x})$ in a posterior for $\evid$ using (H)DPGMM.
\end{enumerate}

In conclusion, the approach presented in this section radically departs from most of the currently adopted numerical schemes, both the ones relying on the stochastic exploration of the parameter space (thermodynamical integration and NS) and the ones using the importance sampling approach: the scheme we propose is able to use the information already contained in the posterior samples and in the corresponding likelihood evaluations to infer the evidence $\evid$, rather than computing its value.

\section{Examples}\label{sec:examples}

In this section we present four different examples in which the evidence is known either analytically or numerically: we apply our inference scheme to each of them to demonstrate that the inferred value of $\evid$ is consistent with the true value $\evid_\mathrm{true}$. Unless otherwise noted, we make use of the NS implementation \textsc{RayNest}\footnote{Publicly available at \url{http://github.com/wdpozzo/raynest}.} to compute the evidence (in all those cases where the integral in Eq.~\ref{eq:evidence} cannot be carried out analytically) and to draw samples from the various distributions. The \textsc{Python} code we developed to perform the inferences presented in both this and the following section is available at \url{https://github.com/sterinaldi/evidence_pop}.

\subsection{Univariate Gaussian distribution}\label{sec:neal}
This first example was presented in \citet{neal:1994} as a simple case in which the harmonic mean estimator in its initial formulation -- Eq.~\eqref{eq:harmonicmean} -- fails. We want to infer the mean $t$ of a Gaussian distribution with known standard deviation $\sigma_1 = 1$ using a single sample $s = 2$ drawn from that distribution. The prior for $t$ is chosen to be Gaussian with $\mu = 0$ and standard deviation $\sigma_0 = 10$. The posterior distribution for $t$ conditioned on $s$ reads:
\begin{equation}
\post{t|s} \propto \like{s|t}\prior{t} = \mathcal{N}(s|t,\sigma_1)\mathcal{N}(t|\mu,\sigma_0)\propto \mathcal{N}\qty(t\Bigg{|}\frac{s/\sigma_1^2}{1/\sigma_1^2+1/\sigma_0^2},\frac{1}{1/\sigma_1^2+1/\sigma_0^2})\,.
\end{equation}

\begin{figure}
\centering
\includegraphics[width = 0.65\columnwidth]{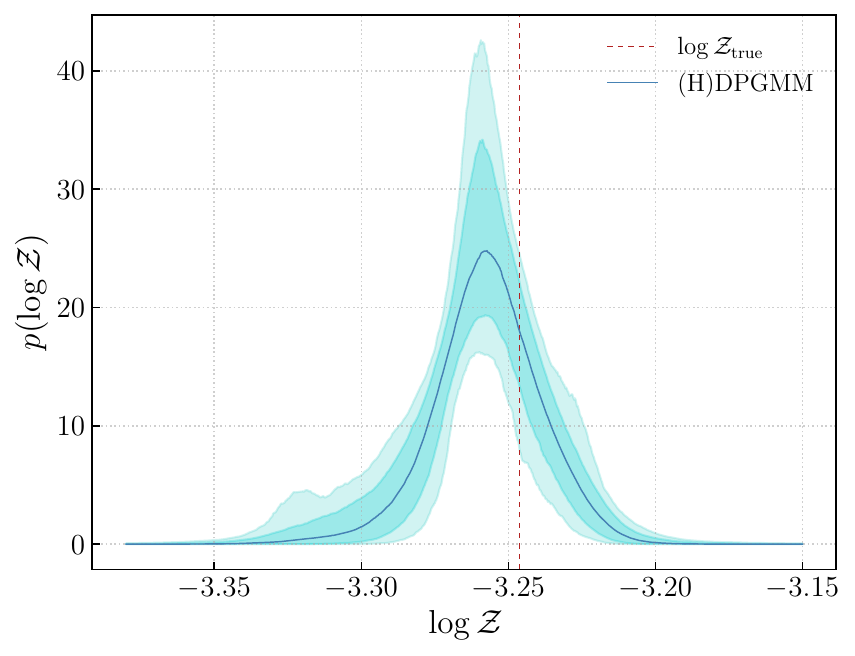}
\caption{Evidence posterior probability density for the univariate Gaussian distribution (Section~\ref{sec:neal}). The dashed red line marks the analytical value for $\log\evid$.}
\label{fig:log_evidence_neal}
\end{figure}

We used a set of 3000 samples from the posterior distribution to reconstruct the DPGMM approximant for the posterior probability density and from these we used a subset of 200 samples to reconstruct $p(\log\evid)$, reported in Figure~\ref{fig:log_evidence_neal}, using the method described in Section~\ref{sec:hierinference}. We get a value of $\log\evid = -3.26^{+0.02}_{-0.02}$, consistent with the expected value $\log\evid_\mathrm{true} = -3.246$.

\begin{figure}
\centering
\includegraphics[width = 0.55\columnwidth]{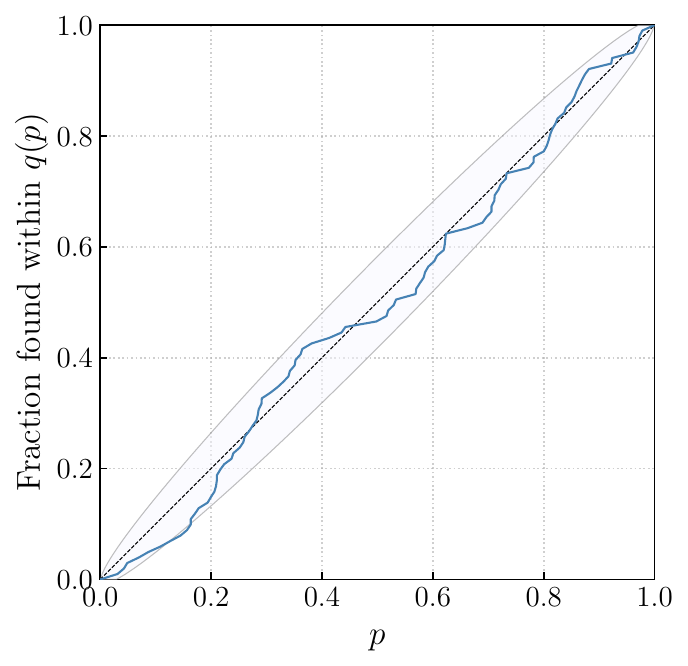}
\caption{Fraction of realisations for which the true evidence $\log\evid_\mathrm{true}$ is found within the quantile $q(p)$ of the median inferred distribution as a function of probability $p$. The shaded region represents the 90\% credible region for the cumulative distribution, estimated from a beta distribution as in \citet{cameron:2011}.}
\label{fig:pp_plot}
\end{figure}

For this specific example we also checked the statistical robustness of our approach making use of the so-called \emph{PP plot test}: we repeated the exercise described in this Section 100 times, each one with a different realisation of the posterior samples: for each realisation we then computed the quantile $q(p)$ of the median inferred distribution at which the true evidence $\log\evid_\mathrm{true}$ is found. The expectation for a robust inference is that $\log\evid_\mathrm{true}$ is found within the quantile $q(p)$ a fraction $p$ of the times. This stems from the fact that the probability of finding the true value at the quantiles $q(p)$ is uniform in the quantiles. The result of this test is reported in Figure~\ref{fig:pp_plot}: the recovered cumulative distribution of quantiles is in agreement with the expectations.

\subsection{Normal-inverse-chi-squared distribution}\label{sec:NIX}
\begin{figure}
\centering
\includegraphics[width = 0.65\columnwidth]{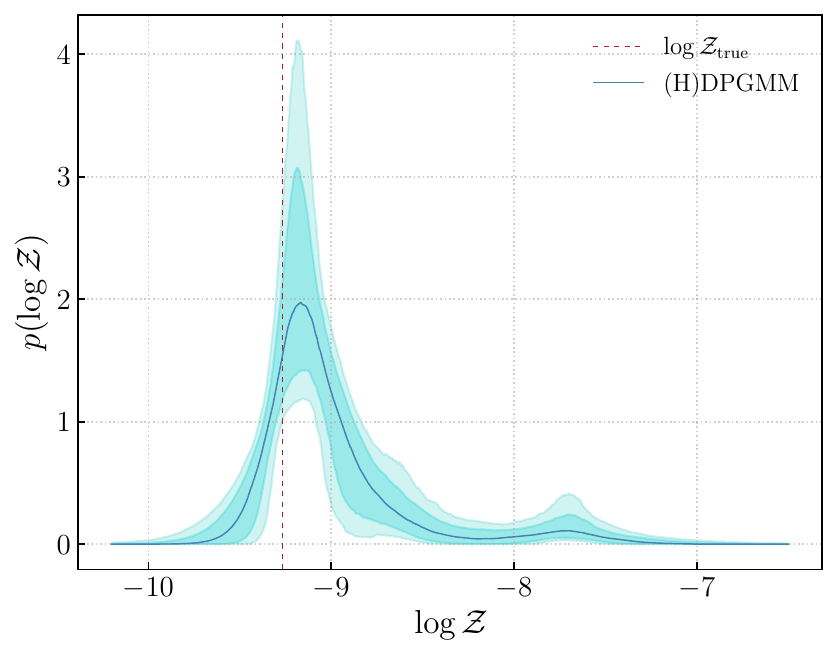}
\caption{Evidence posterior probability density for the $\NIX$ distribution (Section~\ref{sec:NIX}). The dashed red line marks the analytical value for $\log\evid$.}
\label{fig:log_evidence_NIX}
\end{figure}

The second example we give is the normal-inverse-chi-squared ($\NIX$) distribution. This prior for the mean $\mu$ and variance $\sigma^2$ of a Gaussian distribution is conjugated to the Gaussian distribution itself, meaning that under the assumption of a Gaussian likelihood, the posterior distribution will be a $\NIX$ distribution with updated parameters \citep{murphy:2007}. The posterior for $\mu$ and $\sigma^2$ conditioned on two samples $\mathbf{s} = \{-3,7\}$ reads
\begin{multline}
\post{\mu,\sigma^2|\mathbf{s}} \propto \like{\mathbf{s}|\mu,\sigma^2}\prior{\mu,\sigma^2}= \prod_i \mathcal{N}(s_i|\mu,\sigma^2)\NIX(\mu,\sigma^2|\mu_0,\kappa_0,\nu_0,\sigma^2_0)\\ \propto \NIX(\mu,\sigma^2|\mu_n,\kappa_n,\nu_n,\sigma^2_n)\,,
\end{multline}
The $\NIX$ distribution is parametrised as in \citet{murphy:2007} (Eqs. 123-126 and 141-144) with prior parameters $\mu_0 = 0$, $\kappa_0 = 1/10$, $\nu_0 = 1$ and $\sigma_0^2 = 1$. In this case, we used 14050 samples to approximate the posterior distribution $\post{\mu,\sigma^2|\mathbf{s}}$ with the DPGMM and a subset of 1000 samples to infer the evidence. The posterior distribution $p(\evid)$ is reported in Figure~\ref{fig:log_evidence_NIX}: we find $\log\evid = -9.1^{+0.4}_{-0.2}$, consistent with the analytical value of $\log\evid_\mathrm{true} = -9.3$.

\subsection{Parameters of the bivariate Gaussian distribution}\label{sec:bivariate_pars}
Here we consider the case in which we are interested in inferring the 5 parameters of a bivariate Gaussian distribution,
\begin{equation}
\boldsymbol\mu = \mqty(\mu_1\\\mu_2)\qc \boldsymbol \Sigma = \mqty(\sigma_1^2 & \rho_1\sigma_1\sigma_2\\\rho_1\sigma_1\sigma_2&\sigma_2^2)\,,
\end{equation}
using a set of 100 samples $\mathbf{s}=\{s_1,\ldots,s_{100}\}$ from this distribution. In this case, as prior distribution we adopt a uniform distribution $\mathcal{U}(\boldsymbol\mu,\boldsymbol\Sigma)$ within the hypercube defined by the following limits:
\begin{itemize}
\item $\mu_1,\mu_2 \in [-5,5]$;
\item $\sigma_1,\sigma_2 \in [0,10]$;
\item $\rho \in [-1,1]$.
\end{itemize}
The posterior distribution is
\begin{equation}
\post{\boldsymbol\mu,\boldsymbol\Sigma|\mathbf{s}} \propto \like{\mathbf{s}|\boldsymbol\mu,\boldsymbol\Sigma}\prior{\boldsymbol\mu,\boldsymbol\Sigma}\\ = \prod_i \mathcal{N}(s_i|\boldsymbol\mu,\boldsymbol\Sigma)\mathcal{U}(\boldsymbol\mu,\boldsymbol\Sigma)\,.
\end{equation}

\begin{figure}
\centering
\includegraphics[width = 0.65\columnwidth]{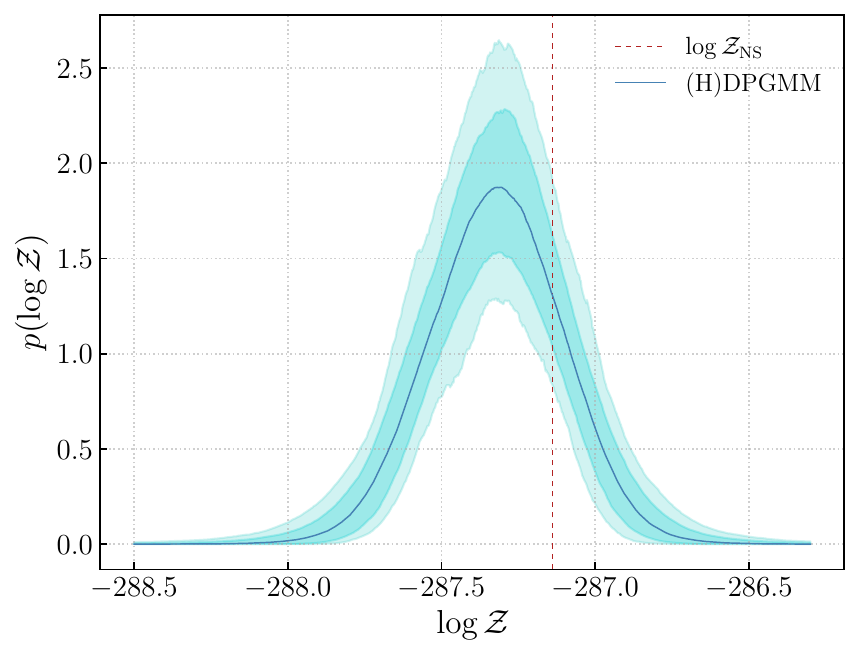}
\caption{Evidence posterior probability density for the parameters of the bivariate Gaussian distribution (Section~\ref{sec:bivariate_pars}). The dashed red line marks the numerical value for $\log\evid$ computed with the NS algorithm.}
\label{fig:log_evidence_bivariate}
\end{figure}

We report the inferred distribution for $\log\evid$ in Figure~\ref{fig:log_evidence_bivariate}. The value we infer with our hierarchical method is $\log\evid = -287.3^{+0.2}_{-0.2}$, consistent with the one obtained via the NS approach, $\log\evid_\mathrm{NS} = -287.1\pm0.1$.

\subsection{GW150914}\label{sec:GW150914}
Our last and most complex example is the evidence for the gravitational-wave (GW) event GW150914 \citep{gw150914:2016,gw150914:properties}, the first GW ever detected. The 15-dimensional parameter space for this problem includes all the parameters required to describe the waveform as observed by the two LIGO detectors \citep{ligodetector:2015}. Details on how the so-called \emph{parameter estimation} (PE) run is performed can be found in \citet{christensen:2022}, being an extensive discussion of the methods beyond the scope of this paper.

The posterior samples of GW150914 are produced both with a Markov Chain Monte-Carlo (MCMC), via the Bilby-MCMC sampler \citep{ashton:2021}, and with a NS scheme, making use of the Dynesty sampler \citep{speagle:2019}. Both these samplers are implemented in the Bilby package \citep{ashton:2019}. The data used in this analysis are publicly available via GWOSC \citep{gwosc:2021}, whereas the settings used for the configuration of the samplers, as they are passed to the \texttt{bilby\_pipe} pipeline \citep{Romero-Shaw:2020}, can be found at \url{https://github.com/sterinaldi/evidence_pop/tree/main/examples/GW/config_files}. A corner plot of the available NS posterior samples is available in the supplementary material.

\begin{figure}
\centering
\includegraphics[width = 0.65\columnwidth]{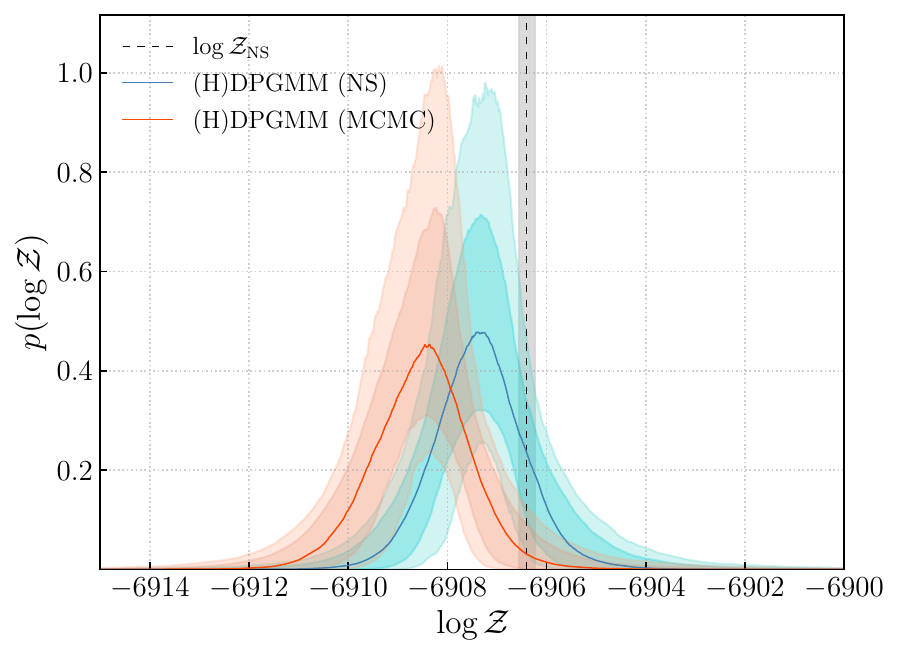}
\caption{Evidence posterior probability density for GW150914 (Section~\ref{sec:GW150914}) obtained using samples from Bilby-MCMC (orange line) and from Dynesty (NS) algorithm (blue line). The dashed black line marks the numerical value for $\log\evid$ computed with the NS algorithm.}
\label{fig:log_evidence_GW150914}
\end{figure}

The recovered evidence distributions are reported in Figure~\ref{fig:log_evidence_GW150914}: using the samples produced with Bilby-MCMC we infer a value of $\log\evid = -6907.3^{+0.9}_{-0.9}$, while using the samples produced via Dynesty we infer a value of $\log\evid = -6908.5^{+0.9}_{-1.0}$, both consistent with the one computed with the NS scheme, $\log\evid_\mathrm{NS} = -6906.4\pm 0.2$.

\section{Bayes factor}\label{sec:bayesfactor}
Within the Bayesian statistics framework, the gold standard to assess whether a model $\hyp{A}$ is favoured by the available data $d$ with respect to a second model $\hyp{B}$ is the so-called \emph{odds ratio}, defined as 
\begin{equation}
\mathcal{O}^\mathrm{A}_\mathrm{B} = \frac{p(\hyp{A}|d)}{p(\hyp{B}|d)} = \frac{p(d|\hyp{A})}{p(d|\hyp{B})}\frac{p(\hyp{A})}{p(\hyp{B})} \equiv \bayesf{A}{B}\mathcal{P}^\mathrm{A}_\mathrm{B}\,.
\end{equation}
The term $\mathcal{P}^\mathrm{A}_\mathrm{B}$, named \emph{prior odds}, conveys the expectation on which of the two models has to be preferred. In general there is no reason to say that one of the model is \emph{a priori} favoured, thus $\mathcal{P}^\mathrm{A}_\mathrm{B}$ is often set equal to one\footnote{There are, nonetheless, situations in which it is reasonable to use a non-trivial prior odds: problems involving rates in astronomy are an example.} and model selection is performed using the \emph{Bayes factor} $\bayesf{A}{B}$. This quantity is defined as the ratio of the evidences of the two competing models:
\begin{equation}
\bayesf{A}{B} \equiv \frac{p(d|\hyp{A})}{p(d|\hyp{B})} = \frac{\evid_\mathrm{A}}{\evid_\mathrm{B}}\,.
\end{equation}
The method introduced in this paper provides a posterior distribution for the evidence of each model, which can be then converted into a posterior distribution for the Bayes' factor:
\begin{multline}
p(\bayesf{A}{B}) = \int p(\bayesf{A}{B}|\evid_\mathrm{A},\evid_\mathrm{B}) p(\evid_\mathrm{A})p(\evid_\mathrm{B})\dd\evid_\mathrm{A}\dd\evid_\mathrm{B}\\ = \int \delta\qty(\bayesf{A}{B} - \frac{\evid_\mathrm{A}}{\evid_\mathrm{B}})p(\evid_\mathrm{A})p(\evid_\mathrm{B})\dd\evid_\mathrm{A}\dd\evid_\mathrm{B}\,.
\end{multline}
We tested this approach in a simulated scenario where we have 100 samples $\mathbf{s}=\{s_1,\ldots,s_{100}\}$ drawn from a Gaussian distribution. The two competing hypothesis are:
\begin{itemize}
\item[$\hyp{N}$:] the samples are drawn from a Gaussian distribution $\mathcal{N}(s|\mu,\sigma)$;
\item[$\hyp{GN}$:] the samples are drawn from a generalised normal distribution $\mathcal{GN}(s|\mu,\alpha,\beta)$.
\end{itemize}
The generalised normal distribution is defined as
\begin{equation}
\mathcal{GN}(s|\mu,\alpha,\beta) = \frac{\beta}{2\alpha\Gamma(1/\beta)}e^{-|s-\mu|^\beta/\alpha}\,.
\end{equation}
The Gaussian distribution is a special case of the generalised normal distribution obtained by setting $\beta = 2$ and $\alpha = \sigma^2/2$, therefore both hypothesis are capable of describing the available data. The Bayes' factor will nonetheless favour the one with less free parameters -- the hypothesis $\hyp{N}$ -- because the additional degree of freedom of the generalised normal distribution does not improve the description of the available data $\mathbf{s}$. We computed the evidence making use of both the method described in this paper and the NS algorithm. For the latter, since it provides a value for $\log\evid$ with an associated uncertainty $\delta\log\evid$, we assumed $p(\log\evid)$ to be a Gaussian distribution.

\begin{figure}
\centering
\includegraphics[width=0.65\columnwidth]{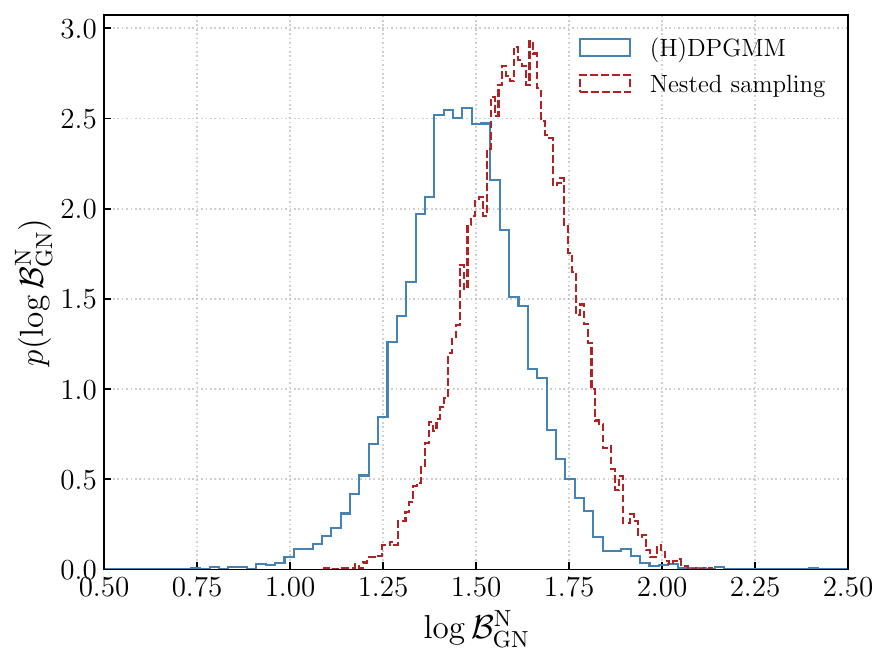}
\caption{$\log\bayesf{N}{GN}$ posterior distribution inferred with our method (solid blue) and computed making use of the NS algorithm (dashed red). The two distributions are consistent with each other.}
\label{fig:logB}
\end{figure}

The two distributions for $p(\log\bayesf{N}{GN})$ are reported in Fig.~\ref{fig:logB}: they are consistent with each other and in agreement with our expectation of the Gaussian distribution to be slightly favoured thanks to the reduced dimensionality of the parameter space.

\section{Summary}\label{sec:summary}

\begin{table}
\caption{Summary of examples presented in Section~\ref{sec:examples}.}
\label{tab:summary}
\begin{center}
\begin{tabular}{lcc}
\toprule
Example & $\log\evid$ & $\log\evid_\mathrm{true/NS}$\\
\midrule
1D Gaussian & $-3.26^{+0.02}_{-0.02}$ & $-3.246$\\
$\NIX$ & $-9.1^{+0.4}_{-0.2}$ & $-9.3$\\
2D Gaussian& $-287.3^{+0.2}_{-0.2}$ & $-287.1\pm 0.1$\\
\midrule
GW150914 &  \begin{tabular}{@{}c@{}}$-6907.3^{+0.9}_{-0.9}$ (NS) \\ $-6908.5^{+0.9}_{-1.0}$ (MCMC)\end{tabular} & $-6906.4\pm 0.2$ \\
\bottomrule
\end{tabular}
\end{center}
\end{table}

In this paper we presented an approach to the evidence computation using posterior samples. Rather than relying on an estimator for $\evid$, we demonstrated that it is possible to use a non-parametric approximant to reconstruct the normalised posterior probability density $\post{x}$ and to obtain a posterior distribution for $\evid$ in a hierarchical fashion.

Since our approach is based on the capability of the employed non-parametric method to faithfully approximate $\post{x}$, it is of the uttermost importance to ensure that this assumption is met. For the non-parametric method used in this work, the DPGMM, this is ensured by the demonstration given in \citet{nguyen:2020}.
Another important thing to keep in mind is that the number of posterior samples used must be large enough to properly represent the underlying posterior distribution (see the discussion in Appendix A of \citet{rinaldi:2022:figaro}).
In order to test our hierarchical scheme, we applied our method to four different scenarios of increasing complexity (summarised in Table~\ref{tab:summary}) and in all cases our hierarchical scheme has been able to produce results that are consistent with the true value of $\log\evid$, when available, or with the one computed using the NS scheme. 

The hierarchical approach presented in this paper can be a viable alternative to the existing methods to compute $\evid$ and will open to the possibility of adopting more efficient stochastic sampling schemes for this task.

\begin{acks}[Acknowledgments]
We thank Stefano~Schmidt for useful comments.

SR is also affiliated to Dipartimento di Fisica e Astronomia ``G. Galilei'', Università di Padova, Via Marzolo 8, 35122 Padova, Italy. GD is also affiliated to INFN, Sezione di Firenze, Via Sansone 1, 50019 Sesto Fiorentino, Italy. WDP is also affilated to INFN, Sezione di Pisa, Largo Bruno Pontecorvo 3, 57127 Pisa, Italy.
\end{acks}
\begin{funding}
SR acknowledges financial support from the European Research Council for the ERC Consolidator grant DEMOBLACK, under contract no. 770017, and from the German Excellence Strategy via the Heidelberg Cluster of Excellence (EXC 2181 - 390900948) STRUCTURES. GD acknowledges financial support from the National Recovery and Resilience Plan, Mission 4 Component 2 – Investment 1.4 – National Center for HPC, Big Data and Quantum Computing – funded by the European Union – NextGenerationEU – CUP (B83C22002830001).

This research made use of the bwForCluster Helix: the authors acknowledge support by the state of Baden-Württemberg through bwHPC and the German Research Foundation (DFG) through grant INST 35/1597-1 FUGG.

This research has made use of data or software obtained from the Gravitational Wave Open Science Center (gwosc.org), a service of the LIGO Scientific Collaboration, the Virgo Collaboration, and KAGRA. This material is based upon work supported by NSF's LIGO Laboratory which is a major facility fully funded by the National Science Foundation, as well as the Science and Technology Facilities Council (STFC) of the United Kingdom, the Max-Planck-Society (MPS), and the State of Niedersachsen/Germany for support of the construction of Advanced LIGO and construction and operation of the GEO600 detector. Additional support for Advanced LIGO was provided by the Australian Research Council. Virgo is funded, through the European Gravitational Observatory (EGO), by the French Centre National de Recherche Scientifique (CNRS), the Italian Istituto Nazionale di Fisica Nucleare (INFN) and the Dutch Nikhef, with contributions by institutions from Belgium, Germany, Greece, Hungary, Ireland, Japan, Monaco, Poland, Portugal, Spain. KAGRA is supported by Ministry of Education, Culture, Sports, Science and Technology (MEXT), Japan Society for the Promotion of Science (JSPS) in Japan; National Research Foundation (NRF) and Ministry of Science and ICT (MSIT) in Korea; Academia Sinica (AS) and National Science and Technology Council (NSTC) in Taiwan.
\end{funding}
\begin{supplement}
\stitle{GW150914 posterior distribution}
\sdescription{Corner plot of the 15-dimensional posterior distribution for GW150914 using posterior samples drawn using the nested sampling scheme.}
\end{supplement}
\bibliographystyle{ba}
\bibliography{bibliography}

\begin{thebibliography}{37}
\newcommand{\enquote}[1]{``#1''}
\expandafter\ifx\csname natexlab\endcsname\relax\def\natexlab#1{#1}\fi
\expandafter\ifx\csname url\endcsname\relax
  \def\url#1{{\tt #1}}\fi
\expandafter\ifx\csname urlprefix\endcsname\relax\def\urlprefix{URL }\fi
\ifx\endbibitem\undefined \let\endbibitem\relax\fi

\bibitem[{{Aasi} et~al.(2015)}]{ligodetector:2015}
{Aasi}, J. et~al. (2015).
\newblock \enquote{{Advanced LIGO}.}
\newblock {\em \cqg\/}, 32: 074001.
\endbibitem

\bibitem[{{Abbott} et~al.(2016{\natexlab{a}}){Abbott}, {Abbott}, {Abbott}, {Abernathy}, {Acernese}, {Ackley}, {Adams}, {Adams}, {Addesso}, {Adhikari}, {Adya} et~al.}]{gw150914:properties}
{Abbott}, B.~P., {Abbott}, R., {Abbott}, T.~D., {Abernathy}, M.~R., {Acernese}, F., {Ackley}, K., {Adams}, C., {Adams}, T., {Addesso}, P., {Adhikari}, R.~X., {Adya}, V.~B., et~al. (2016{\natexlab{a}}).
\newblock \enquote{{Properties of the Binary Black Hole Merger GW150914}.}
\newblock {\em \prl\/}, 116(24): 241102.
\endbibitem

\bibitem[{{Abbott} et~al.(2016{\natexlab{b}})}]{gw150914:2016}
{Abbott}, B.~P. et~al. (2016{\natexlab{b}}).
\newblock \enquote{Observation of Gravitational Waves from a Binary Black Hole Merger.}
\newblock {\em Phys. Rev. Lett.\/}, 116: 061102.
\newline\urlprefix\url{https://link.aps.org/doi/10.1103/PhysRevLett.116.061102}
\endbibitem

\bibitem[{{Abbott} et~al.(2021)}]{gwosc:2021}
{Abbott}, R. et~al. (2021).
\newblock \enquote{{Open data from the first and second observing runs of Advanced LIGO and Advanced Virgo}.}
\newblock {\em SoftwareX\/}, 13: 100658.
\endbibitem

\bibitem[{{Ashton} et~al.(2019){Ashton}, {H{\"u}bner}, {Lasky}, {Talbot}, {Ackley}, {Biscoveanu}, {Chu}, {Divakarla}, {Easter}, {Goncharov}, {Hernandez Vivanco}, {Harms}, {Lower}, {Meadors}, {Melchor}, {Payne}, {Pitkin}, {Powell}, {Sarin}, {Smith}, and {Thrane}}]{ashton:2019}
{Ashton}, G., {H{\"u}bner}, M., {Lasky}, P.~D., {Talbot}, C., {Ackley}, K., {Biscoveanu}, S., {Chu}, Q., {Divakarla}, A., {Easter}, P.~J., {Goncharov}, B., {Hernandez Vivanco}, F., {Harms}, J., {Lower}, M.~E., {Meadors}, G.~D., {Melchor}, D., {Payne}, E., {Pitkin}, M.~D., {Powell}, J., {Sarin}, N., {Smith}, R. J.~E., and {Thrane}, E. (2019).
\newblock \enquote{{BILBY: A User-friendly Bayesian Inference Library for Gravitational-wave Astronomy}.}
\newblock {\em The Astrophysical Journal Supplement Series\/}, 241(2): 27.
\endbibitem

\bibitem[{{Ashton} and {Talbot}(2021)}]{ashton:2021}
{Ashton}, G. and {Talbot}, C. (2021).
\newblock \enquote{{BILBY-MCMC: an MCMC sampler for gravitational-wave inference}.}
\newblock {\em Monthly Notices of the Royal Astronomical Society\/}, 507(2): 2037--2051.
\endbibitem

\bibitem[{{Caldwell} et~al.(2018){Caldwell}, {Eller}, {Hafych}, {Schick}, {Schulz}, and {Szalay}}]{caldwell:2018}
{Caldwell}, A., {Eller}, P., {Hafych}, V., {Schick}, R.~C., {Schulz}, O., and {Szalay}, M. (2018).
\newblock \enquote{{Integration with an Adaptive Harmonic Mean Algorithm}.}
\newblock {\em arXiv e-prints\/}, arXiv:1808.08051.
\endbibitem

\bibitem[{Cameron(2011)}]{cameron:2011}
Cameron, E. (2011).
\newblock \enquote{On the Estimation of Confidence Intervals for Binomial Population Proportions in Astronomy: The Simplicity and Superiority of the Bayesian Approach.}
\newblock {\em Publications of the Astronomical Society of Australia\/}, 28(2): 128–139.
\endbibitem

\bibitem[{{Chib}(1995)}]{chib:1995}
{Chib}, S. (1995).
\newblock \enquote{Marginal Likelihood from the Gibbs Output.}
\newblock {\em Journal of the American Statistical Association\/}, 90(432): 1313--1321.
\newline\urlprefix\url{http://www.jstor.org/stable/2291521}
\endbibitem

\bibitem[{{Christensen} and {Meyer}(2022)}]{christensen:2022}
{Christensen}, N. and {Meyer}, R. (2022).
\newblock \enquote{{Parameter estimation with gravitational waves}.}
\newblock {\em Reviews of Modern Physics\/}, 94(2): 025001.
\endbibitem

\bibitem[{{Clyde} et~al.(2007){Clyde}, {Berger}, {Bullard}, {Ford}, {Jefferys}, {Luo}, {Paulo}, and {Loredo}}]{clyde:2007}
{Clyde}, M.~A., {Berger}, J.~O., {Bullard}, F., {Ford}, E.~B., {Jefferys}, W.~H., {Luo}, R., {Paulo}, R., and {Loredo}, T. (2007).
\newblock \enquote{{Current Challenges in Bayesian Model Choice}.}
\newblock In {Babu}, G.~J. and {Feigelson}, E.~D. (eds.), {\em Statistical Challenges in Modern Astronomy IV\/}, volume 371 of {\em Astronomical Society of the Pacific Conference Series\/}, 224.
\endbibitem

\bibitem[{{Escobar} and {West}(1995)}]{escobar:1995}
{Escobar}, M.~D. and {West}, M. (1995).
\newblock \enquote{Bayesian Density Estimation and Inference Using Mixtures.}
\newblock {\em Journal of the American Statistical Association\/}, 90(430): 577--588.
\newline\urlprefix\url{http://www.jstor.org/stable/2291069}
\endbibitem

\bibitem[{{Friel} and {Wyse}(2012)}]{friel:2012}
{Friel}, N. and {Wyse}, J. (2012).
\newblock \enquote{Estimating the evidence – a review.}
\newblock {\em Statistica Neerlandica\/}, 66(3): 288--308.
\newline\urlprefix\url{https://onlinelibrary.wiley.com/doi/abs/10.1111/j.1467-9574.2011.00515.x}
\endbibitem

\bibitem[{{Gelfand} and {Dey}(1994)}]{gelfand:1994}
{Gelfand}, A.~E. and {Dey}, D.~K. (1994).
\newblock \enquote{Bayesian Model Choice: Asymptotics and Exact Calculations.}
\newblock {\em Journal of the Royal Statistical Society. Series B (Methodological)\/}, 56(3): 501--514.
\newline\urlprefix\url{http://www.jstor.org/stable/2346123}
\endbibitem

\bibitem[{{Gelman} et~al.(2013){Gelman}, {Carlin}, {Stern}, {Dunson}, {Vehtari}, and {Rubin}}]{gelman:2013}
{Gelman}, A., {Carlin}, J., {Stern}, H., {Dunson}, D., {Vehtari}, A., and {Rubin}, D. (2013).
\newblock {\em Bayesian Data Analysis, Third Edition\/}.
\newblock Chapman \& Hall/CRC Texts in Statistical Science. Taylor \& Francis.
\newline\urlprefix\url{https://books.google.it/books?id=ZXL6AQAAQBAJ}
\endbibitem

\bibitem[{{Gelman} and {Meng}(1998)}]{gelman:1998}
{Gelman}, A. and {Meng}, X.-L. (1998).
\newblock \enquote{Simulating normalizing constants: from importance sampling to bridge sampling to path sampling.}
\newblock {\em Statistical Science\/}, 13(2): 163 -- 185.
\newline\urlprefix\url{https://doi.org/10.1214/ss/1028905934}
\endbibitem

\bibitem[{{Hobson} and {McLachlan}(2003)}]{hobson:2003}
{Hobson}, M.~P. and {McLachlan}, C. (2003).
\newblock \enquote{{A Bayesian approach to discrete object detection in astronomical data sets}.}
\newblock {\em \mnras\/}, 338(3): 765--784.
\endbibitem

\bibitem[{{Lee} and {Robert}(2016)}]{lee:2016}
{Lee}, J.~E. and {Robert}, C.~P. (2016).
\newblock \enquote{{Importance Sampling Schemes for Evidence Approximation in Mixture Models}.}
\newblock {\em Bayesian Analysis\/}, 11(2): 573 -- 597.
\newline\urlprefix\url{https://doi.org/10.1214/15-BA970}
\endbibitem

\bibitem[{{Liu}(1994)}]{liu:1994}
{Liu}, J.~S. (1994).
\newblock \enquote{The Collapsed Gibbs Sampler in Bayesian Computations with Applications to a Gene Regulation Problem.}
\newblock {\em Journal of the American Statistical Association\/}, 89(427): 958--966.
\newline\urlprefix\url{http://www.jstor.org/stable/2290921}
\endbibitem

\bibitem[{{McEwen} et~al.(2021){McEwen}, {Wallis}, {Price}, and {Spurio Mancini}}]{mcewen:2021}
{McEwen}, J.~D., {Wallis}, C. G.~R., {Price}, M.~A., and {Spurio Mancini}, A. (2021).
\newblock \enquote{{Machine learning assisted Bayesian model comparison: learnt harmonic mean estimator}.}
\newblock {\em arXiv e-prints\/}, arXiv:2111.12720.
\endbibitem

\bibitem[{Meng and Wong(1996)}]{meng:1996}
Meng, X.-L. and Wong, W.~H. (1996).
\newblock \enquote{Simulating ratios of normalizing constants via a simple identity: a theoretical exploration.}
\newblock {\em Statist. Sinica\/}, 6(4): 831--860.
\endbibitem

\bibitem[{{Murphy}(2007)}]{murphy:2007}
{Murphy}, K.~P. (2007).
\newblock \enquote{Conjugate Bayesian analysis of the Gaussian distribution.}
\newline\urlprefix\url{https://www.cs.ubc.ca/~murphyk/Papers/bayesGauss.pdf}
\endbibitem

\bibitem[{{Neal}(1994)}]{neal:1994}
{Neal}, R. (1994).
\newblock \enquote{The Harmonic Mean of the Likelihood: Worst Monte Carlo Method Ever.}
\newline\urlprefix\url{https://radfordneal.wordpress.com/2008/08/17/the-harmonic-mean-of-the-likelihood-worst-monte-carlo-method-ever/}
\endbibitem

\bibitem[{{Newton} and {Raftery}(1994)}]{newton:1994}
{Newton}, M.~A. and {Raftery}, A.~E. (1994).
\newblock \enquote{Approximate Bayesian Inference with the Weighted Likelihood Bootstrap.}
\newblock {\em Journal of the Royal Statistical Society. Series B (Methodological)\/}, 56(1): 3--48.
\newline\urlprefix\url{http://www.jstor.org/stable/2346025}
\endbibitem

\bibitem[{{Nguyen} et~al.(2020){Nguyen}, {Nguyen}, {Chamroukhi}, and {McLachlan}}]{nguyen:2020}
{Nguyen}, T.~T., {Nguyen}, H.~D., {Chamroukhi}, F., and {McLachlan}, G.~J. (2020).
\newblock \enquote{{Approximation by finite mixtures of continuous density functions that vanish at infinity}.}
\newblock {\em Cog. Math. St.\/}, 7(1): 1750861.
\newline\urlprefix\url{https://doi.org/10.1080/25742558.2020.1750861}
\endbibitem

\bibitem[{{Polanska} et~al.(2023){Polanska}, {Price}, {Spurio Mancini}, and {McEwen}}]{polanska:2023}
{Polanska}, A., {Price}, M.~A., {Spurio Mancini}, A., and {McEwen}, J.~D. (2023).
\newblock \enquote{{Learned harmonic mean estimation of the marginal likelihood with normalizing flows}.}
\newblock {\em arXiv e-prints\/}, arXiv:2307.00048.
\endbibitem

\bibitem[{{Rasmussen}(2000)}]{rasmussen:2000}
{Rasmussen}, C. (2000).
\newblock \enquote{The Infinite Gaussian Mixture Model.}
\newblock In {Solla}, S., {Leen}, T., and {M\"{u}ller}, K. (eds.), {\em Advances in Neural Information Processing Systems\/}, volume~12. MIT Press.
\newline\urlprefix\url{https://proceedings.neurips.cc/paper/1999/file/97d98119037c5b8a9663cb21fb8ebf47-Paper.pdf}
\endbibitem

\bibitem[{{Rinaldi} and {Del Pozzo}(2022{\natexlab{a}})}]{rinaldi:2022:hdpgmm}
{Rinaldi}, S. and {Del Pozzo}, W. (2022{\natexlab{a}}).
\newblock \enquote{{(H)DPGMM: a hierarchy of Dirichlet process Gaussian mixture models for the inference of the black hole mass function}.}
\newblock {\em \mnras\/}, 509(4): 5454--5466.
\endbibitem

\bibitem[{{Rinaldi} and {Del Pozzo}(2022{\natexlab{b}})}]{rinaldi:2022:figaro}
--- (2022{\natexlab{b}}).
\newblock \enquote{{Rapid localization of gravitational wave hosts with FIGARO}.}
\newblock {\em \mnras\/}, 517(1): L5--L10.
\endbibitem

\bibitem[{{Robert} and {Wraith}(2009)}]{robert:2009}
{Robert}, C.~P. and {Wraith}, D. (2009).
\newblock \enquote{{Computational methods for Bayesian model choice}.}
\newblock In {Goggans}, P.~M. and {Chan}, C.-Y. (eds.), {\em Bayesian Inference and Maximum Entropy Methods in Science and Engineering: The 29th International Workshop on Bayesian Inference and Maximum Entropy Methods in Science and Engineering\/}, volume 1193 of {\em American Institute of Physics Conference Series\/}, 251--262.
\endbibitem

\bibitem[{{Romero-Shaw} et~al.(2020){Romero-Shaw}, {Talbot}, {Biscoveanu}, {D'Emilio}, {Ashton}, {Berry}, {Coughlin}, {Galaudage}, {Hoy}, {H{\"u}bner}, {Phukon}, {Pitkin}, {Rizzo}, {Sarin}, {Smith}, {Stevenson}, {Vajpeyi}, {Ar{\`e}ne}, {Athar}, {Banagiri}, {Bose}, {Carney}, {Chatziioannou}, {Clark}, {Colleoni}, {Cotesta}, {Edelman}, {Estell{\'e}s}, {Garc{\'\i}a-Quir{\'o}s}, {Ghosh}, {Green}, {Haster}, {Husa}, {Keitel}, {Kim}, {Hernandez-Vivanco}, {Maga{\~n}a Hernandez}, {Karathanasis}, {Lasky}, {De Lillo}, {Lower}, {Macleod}, {Mateu-Lucena}, {Miller}, {Millhouse}, {Morisaki}, {Oh}, {Ossokine}, {Payne}, {Powell}, {Pratten}, {P{\"u}rrer}, {Ramos-Buades}, {Raymond}, {Thrane}, {Veitch}, {Williams}, {Williams}, and {Xiao}}]{Romero-Shaw:2020}
{Romero-Shaw}, I.~M., {Talbot}, C., {Biscoveanu}, S., {D'Emilio}, V., {Ashton}, G., {Berry}, C.~P.~L., {Coughlin}, S., {Galaudage}, S., {Hoy}, C., {H{\"u}bner}, M., {Phukon}, K.~S., {Pitkin}, M., {Rizzo}, M., {Sarin}, N., {Smith}, R., {Stevenson}, S., {Vajpeyi}, A., {Ar{\`e}ne}, M., {Athar}, K., {Banagiri}, S., {Bose}, N., {Carney}, M., {Chatziioannou}, K., {Clark}, J.~A., {Colleoni}, M., {Cotesta}, R., {Edelman}, B., {Estell{\'e}s}, H., {Garc{\'\i}a-Quir{\'o}s}, C., {Ghosh}, A., {Green}, R., {Haster}, C.~J., {Husa}, S., {Keitel}, D., {Kim}, A.~X., {Hernandez-Vivanco}, F., {Maga{\~n}a Hernandez}, I., {Karathanasis}, C., {Lasky}, P.~D., {De Lillo}, N., {Lower}, M.~E., {Macleod}, D., {Mateu-Lucena}, M., {Miller}, A., {Millhouse}, M., {Morisaki}, S., {Oh}, S.~H., {Ossokine}, S., {Payne}, E., {Powell}, J., {Pratten}, G., {P{\"u}rrer}, M., {Ramos-Buades}, A., {Raymond}, V., {Thrane}, E., {Veitch}, J., {Williams}, D., {Williams}, M.~J., and {Xiao}, L. (2020).
\newblock \enquote{{Bayesian inference for compact binary coalescences with BILBY: validation and application to the first LIGO-Virgo gravitational-wave transient catalogue}.}
\newblock {\em \mnras\/}, 499(3): 3295--3319.
\endbibitem

\bibitem[{{Sivia} and {Skilling}(2006)}]{sivia:2006}
{Sivia}, D.~S. and {Skilling}, J. (2006).
\newblock {\em {Data Analysis - A Bayesian Tutorial}\/}.
\newblock Oxford Science Publications. Oxford University Press, 2nd edition.
\endbibitem

\bibitem[{{Skilling}(2006)}]{skilling:2006}
{Skilling}, J. (2006).
\newblock \enquote{{Nested sampling for general Bayesian computation}.}
\newblock {\em Bayesian Analysis\/}, 1(4): 833 -- 859.
\newline\urlprefix\url{https://doi.org/10.1214/06-BA127}
\endbibitem

\bibitem[{{Speagle}(2020)}]{speagle:2019}
{Speagle}, J.~S. (2020).
\newblock \enquote{{DYNESTY: a dynamic nested sampling package for estimating Bayesian posteriors and evidences}.}
\newblock {\em Monthly Notices of the Royal Astronomical Society\/}, 493(3): 3132--3158.
\endbibitem

\bibitem[{{Spurio Mancini} et~al.(2023){Spurio Mancini}, {Docherty}, {Price}, and {McEwen}}]{spuriomancini:2023}
{Spurio Mancini}, A., {Docherty}, M.~M., {Price}, M.~A., and {McEwen}, J.~D. (2023).
\newblock \enquote{{Bayesian model comparison for simulation-based inference}.}
\newblock {\em RAS Techniques and Instruments\/}, 2(1): 710--722.
\endbibitem

\bibitem[{{Srinivasan} et~al.(2024){Srinivasan}, {Crisostomi}, {Trotta}, {Barausse}, and {Breschi}}]{srinivasan:2024}
{Srinivasan}, R., {Crisostomi}, M., {Trotta}, R., {Barausse}, E., and {Breschi}, M. (2024).
\newblock \enquote{{floZ: Evidence estimation from posterior samples with normalizing flows}.}
\newblock {\em arXiv e-prints\/}, arXiv:2404.12294.
\endbibitem

\bibitem[{{Tierney} and {Kadane}(1986)}]{tierney:1986}
{Tierney}, L. and {Kadane}, J.~B. (1986).
\newblock \enquote{Accurate Approximations for Posterior Moments and Marginal Densities.}
\newblock {\em Journal of the American Statistical Association\/}, 81(393): 82--86.
\newline\urlprefix\url{http://www.jstor.org/stable/2287970}
\endbibitem

\end{thebibliography}
\end{document}